\documentclass[twocolumn,tighten,times, trackchanges]{aastex7}

\newcommand{\mbh}{\ifmmode {M_{\bullet}} \else $M_{\bullet}$\fi}
\newcommand{\msun}{\ifmmode {M_{\odot}} \else $M_{\odot}$\fi}
\def\ergscma{\rm erg\,s^{-1}\,cm^{-2}\,\AA^{-1}}

\begin{document}

\title{Discovery of a Luminosity-dependent Continuum Lag in NGC 4151 from Photometric and Spectroscopic Continuum Reverberation Mapping}
\shorttitle{Continuum Reverberation Mapping of NGC 4151}

\shortauthors{Feng et al.}

\author[0000-0002-1530-2680]{Hai-Cheng Feng}
\affiliation{Yunnan Observatories, Chinese Academy of Sciences, Kunming 650216, Yunnan, People's Republic of China}
\affiliation{Key Laboratory for the Structure and Evolution of Celestial Objects, Chinese Academy of Sciences, Kunming 650216, Yunnan, People's Republic of China}
\affiliation{Center for Astronomical Mega-Science, Chinese Academy of Sciences, 20A Datun Road, Chaoyang District, Beijing 100012, People's Republic of China}
\affiliation{Key Laboratory of Radio Astronomy and Technology, Chinese Academy of Sciences, 20A Datun Road, Chaoyang District, Beijing 100101, People's Republic of China}
\email[show]{hcfeng@ynao.ac.cn}

\author[0000-0003-3823-3419]{Sha-Sha Li}
\affiliation{Yunnan Observatories, Chinese Academy of Sciences, Kunming 650216, Yunnan, People's Republic of China}
\affiliation{Key Laboratory for the Structure and Evolution of Celestial Objects, Chinese Academy of Sciences, Kunming 650216, Yunnan, People's Republic of China}
\affiliation{Center for Astronomical Mega-Science, Chinese Academy of Sciences, 20A Datun Road, Chaoyang District, Beijing 100012, People's Republic of China}
\affiliation{Key Laboratory of Radio Astronomy and Technology, Chinese Academy of Sciences, 20A Datun Road, Chaoyang District, Beijing 100101, People's Republic of China}
\email[show]{lishasha@ynao.ac.cn}

\author[0000-0002-0771-2153]{Mouyuan Sun}
\affiliation{Department of Astronomy, Xiamen University, Xiamen, Fujian 361005, People’s Republic of China}
\email[show]{msun88@xmu.edu.cn}

\author[0000-0003-2532-7379]{Ciro Pinto}
\affiliation{INAF--IASF Palermo, Via U. La Malfa 153, I-90146 Palermo, Italy}
\email{}

\author[0009-0005-2801-6594]{Shuying Zhou}
\affiliation{Department of Astronomy, Xiamen University, Xiamen, Fujian 361005, People’s Republic of China}
\email{}

\author[0000-0002-2523-5485]{Yerong Xu}
\affiliation{Institute of Space Sciences (ICE, CSIC), Campus UAB, Carrer de Can Magrans s/n, 08193, Barcelona, Spain}
\affiliation{Institut d'Estudis Espacials de Catalunya (IEEC), Esteve Terradas 1, RDIT Building, Of. 212 Mediterranean Technology Park (PMT), 08860, Castelldefels, Spain}
\email{}

\author{J. M. Bai}
\affiliation{Yunnan Observatories, Chinese Academy of Sciences, Kunming 650216, Yunnan, People's Republic of China}
\affiliation{Key Laboratory for the Structure and Evolution of Celestial Objects, Chinese Academy of Sciences, Kunming 650216, Yunnan, People's Republic of China}
\affiliation{Center for Astronomical Mega-Science, Chinese Academy of Sciences, 20A Datun Road, Chaoyang District, Beijing 100012, People's Republic of China}
\affiliation{Key Laboratory of Radio Astronomy and Technology, Chinese Academy of Sciences, 20A Datun Road, Chaoyang District, Beijing 100101, People's Republic of China}
\email{}

\author[0000-0001-9931-8681]{Elena Dalla Bont\`a}
\affiliation{Dipartimento di Fisica e Astronomia ``G. Galilei”, Universit\`a di Padova, Vicolo dell’Osservatorio 3, I-35122 Padova, Italy}
\affiliation{INAF--Osservatorio Astronomico di Padova, Vicolo dell'Osservatorio 5, I-35122 Padova, Italy}
\affiliation{Jeremiah Horrocks Institute, University of Central Lancashire, Preston, PR1 2HE, UK}
\email{}

\author[0009-0003-5592-3734]{ZhongNan Dong}
\affiliation{School of Physics and Astronomy, Sun Yat-sen University, Zhuhai 519082, People's Republic of China}
\affiliation{CSST Science Center for the Guangdong-Hong Kong-Macau Greater Bay Area, Zhuhai 519082, People's Republic of China}
\affiliation{National Astronomical Observatories, Chinese Academy of Sciences, Beĳing 100101, People's Republic of China}
\email{}

\author[0000-0003-0071-8947]{Neeraj Kumari}
\affiliation{INAF--IASF Palermo, Via U. La Malfa 153, I-90146 Palermo, Italy}
\email{}

\author[0000-0002-3134-9526]{Jiaqi Lin}
\affiliation{School of Physics and Astronomy, Sun Yat-sen University, Zhuhai 519082, People's Republic of China}
\affiliation{CSST Science Center for the Guangdong-Hong Kong-Macau Greater Bay Area, Zhuhai 519082, People's Republic of China}
\affiliation{Shanghai Astronomical Observatory, Chinese Academy of Sciences, 80 Nandan Road, Shanghai, 200030, People's Republic of China}
\email{linjq63@mail2.sysu.edu.cn}

\author[0000-0002-2153-3688]{H. T. Liu}
\affiliation{Yunnan Observatories, Chinese Academy of Sciences, Kunming 650216, Yunnan, People's Republic of China}
\affiliation{Key Laboratory for the Structure and Evolution of Celestial Objects, Chinese Academy of Sciences, Kunming 650216, Yunnan, People's Republic of China}
\affiliation{Center for Astronomical Mega-Science, Chinese Academy of Sciences, 20A Datun Road, Chaoyang District, Beijing 100012, People's Republic of China}
\email{}

\author[0000-0002-2310-0982]{Kai-Xing Lu}
\affiliation{Yunnan Observatories, Chinese Academy of Sciences, Kunming 650216, Yunnan, People's Republic of China}
\affiliation{Key Laboratory for the Structure and Evolution of Celestial Objects, Chinese Academy of Sciences, Kunming 650216, Yunnan, People's Republic of China}
\affiliation{Center for Astronomical Mega-Science, Chinese Academy of Sciences, 20A Datun Road, Chaoyang District, Beijing 100012, People's Republic of China}
\email{}

\author[0000-0002-6077-6287]{Bin Ma} 
\affiliation{School of Physics and Astronomy, Sun Yat-sen University, Zhuhai 519082, People's Republic of China}
\affiliation{CSST Science Center for the Guangdong-Hong Kong-Macau Greater Bay Area, Zhuhai 519082, People's Republic of China}
\email{}

\author[0000-0002-7077-7195]{Ji-Rong Mao}
\affiliation{Yunnan Observatories, Chinese Academy of Sciences, Kunming 650216, Yunnan, People's Republic of China}
\affiliation{Key Laboratory for the Structure and Evolution of Celestial Objects, Chinese Academy of Sciences, Kunming 650216, Yunnan, People's Republic of China}
\affiliation{Center for Astronomical Mega-Science, Chinese Academy of Sciences, 20A Datun Road, Chaoyang District, Beijing 100012, People's Republic of China}
\email{}

\author[0000-0001-9226-8992]{Emanuele Nardini}
\affiliation{INAF--Osservatorio Astrofisico di Arcetri, Largo E. Fermi 5, 50125 Firenze,Italy}
\email{}

\author[0000-0001-9095-2782]{Enrico Piconcelli}
\affiliation{INAF--Osservatorio Astronomico di Roma, Via Frascati 33, 00040 Monte Porzio Catone, Italy}
\email{}

\author[0000-0002-3869-2925]{Fabio Pintore}
\affiliation{INAF--IASF Palermo, Via U. La Malfa 153, 90146 Palermo, Italy}
\email{}

\author[0000-0003-4156-3793]{Jian-Guo Wang}
\affiliation{Yunnan Observatories, Chinese Academy of Sciences, Kunming 650216, Yunnan, People's Republic of China}
\affiliation{Key Laboratory for the Structure and Evolution of Celestial Objects, Chinese Academy of Sciences, Kunming 650216, Yunnan, People's Republic of China}
\affiliation{Center for Astronomical Mega-Science, Chinese Academy of Sciences, 20A Datun Road, Chaoyang District, Beijing 100012, People's Republic of China}
\email{}

\author[0000-0003-4156-3793]{Ding-Rong Xiong}
\affiliation{Yunnan Observatories, Chinese Academy of Sciences, Kunming 650216, Yunnan, People's Republic of China}
\affiliation{Key Laboratory for the Structure and Evolution of Celestial Objects, Chinese Academy of Sciences, Kunming 650216, Yunnan, People's Republic of China}
\affiliation{Center for Astronomical Mega-Science, Chinese Academy of Sciences, 20A Datun Road, Chaoyang District, Beijing 100012, People's Republic of China}
\email{}


\begin{abstract}
Accretion onto supermassive black holes (SMBHs) powers active galactic nuclei (AGNs) and drives feedback that shapes galaxy evolution. Constraining AGN accretion disk structure is therefore essential for understanding black hole growth and feedback processes. However, direct constraints on disk size remain rare---particularly from long-term, multi-season spectroscopic reverberation mapping (RM), which is critical for isolating the intrinsic disk response from the broad-line region (BLR). We present results from an intensive multi-wavelength RM campaign of NGC 4151 during its brightest state in nearly two decades. This represents the third high-cadence monitoring over the past decade, capturing accretion states spanning the transitional regime between thin and thick disks, making NGC 4151 the only AGN with continuum RM observations across such a wide range in accretion states. Combining spectroscopy from the Lijiang 2.4 m telescope with coordinated Swift UV/X-ray monitoring, we measure inter-band continuum lags from UV to optical. The wavelength-dependent lags follow a tight $\tau \propto \lambda^{4/3}$ relation, consistent with reprocessing in a thin disk, but exceed theoretical predictions by a factor of 6.6. Our lag spectrum reveals clear excesses near the Balmer and possibly Paschen jumps, confirming diffuse continuum (DC) contamination from the BLR. By comparing the three campaigns, we discover a non-monotonic lag-luminosity trend ($>3\sigma$), which cannot be explained by DC emission alone. We propose the lags reflect combined disk and BLR contributions, and present the first evidence that the DC component follows an intrinsic Baldwin effect. These results offer new insights into SMBH mass measurements and theoretical models of AGN inner structure.
\end{abstract}



\section{Introduction} \label{sec:1}
Over the past two decades, numerous observations have revealed tight correlations between supermassive black hole (SMBH) mass and host galaxy properties---such as bulge stellar velocity dispersion and luminosity \citep{Magorrian1998, Gebhardt2000}. These correlations span more than four orders of magnitude in SMBH mass, suggesting that SMBHs and their host galaxies might have undergone co-evolution \citep{Kormendy2013}. Understanding the physical mechanisms driving this co-evolution requires observational constraints in order to unravel both the growth of SMBHs and their interactions with the surrounding galactic environment.

Rapid accretion of material, which powers active galactic nuclei (AGNs), is considered one of the primary channels for SMBH growth and galaxy feedback \citep[e.g.,][]{Inayoshi2020}. During this process, infalling material forms an accretion disk that efficiently converts gravitational energy into radiation \citep{Rees1984}. This energy release drives feedback through radiation pressure or outflows that regulate star formation across the galaxy \citep[see][for a review]{Fabian2012}. Theoretical models predict that the Eddington ratio fundamentally determines the disk structure \citep{Shakura1973, Narayan1995, Abramowicz1988, Esin1997}. Therefore, characterizing the structure of AGN accretion disks with various Eddington ratios is essential for revealing SMBH growth and feedback in galaxy evolution.

Direct observations of structural evolution in AGN accretion disks remain observationally challenging. For a $10^8 M_\odot$ SMBH, the disk structure evolves through viscous processes operating on timescales of $\sim10^5$ years \citep[e.g.,][]{Noda2018}, far exceeding practical monitoring durations. A promising breakthrough comes from the discovery of changing-look (CL) AGN, a rare class of sources that exhibit extreme variations in luminosity along with the appearance/disappearance of broad emission lines \citep[e.g.,][]{Shappee2014, Guo2024, Dong2025}. These dramatic transitions are widely interpreted as signatures of rapid changes in the accretion rate \citep{Sheng2017, Feng2021, Feng2024, Ren2024a}, although the underlying physical mechanisms remain debated \citep{Ross2018, Sniegowska2020}. Monitoring CL AGN across different phases, we have the opportunity to investigate how the disk structure responds to variations of the accretion rate.

Continuum reverberation mapping (RM) has emerged as a powerful tool to probe the physical scale of AGN accretion disks. In the framework of the X-ray reprocessing model, high-energy photons from the corona irradiate the accretion disk, where they are absorbed and re-emitted at longer wavelengths. Since shorter-wavelength emission arises from the hotter, inner regions of the disk and longer-wavelength emission from the cooler, outer regions, light-travel time differences produce wavelength-dependent lags that scale as $\tau \propto \lambda^{4/3}$ in the standard thin-disk model \citep{Cackett2007}. Intensive monitoring of a dozen AGNs has confirmed this relation. However, the observed lags are often several times larger than those predicted within the X-ray reprocessing scenario \citep[e.g.,][]{McHardy2014, Edelson2017, Guo2022, Gonzalez-Buitrago2023, Gonzalez-Buitrago2025, Kara2023}. 

Two categories of explanations have been proposed to reconcile this discrepancy. One attempts to modify the theory of standard disks, e.g., by considering non-blackbody emission from scattering effects \citep{Hall2018} or magnetic coupling between corona and disk \citep{Sun2020}. The alternative scenario instead invokes contamination from reprocessed emission on larger scales, particularly the broad emission lines and diffuse continuum (DC) from the broad-line region (BLR) \citep{Korista2019}. Most previous continuum RM studies \citep[except for NGC 4593 by][]{Cackett2018} rely on broadband photometric monitoring, which inevitably suffers from contamination by broad emission lines. Moreover, although a lag excess is frequently seen in the $U$/$u$ band---where the Balmer continuum contributes significantly---this feature is limited to a single photometric band, providing minimal constraints on the wavelength-dependent behavior of the DC component.

These challenges motivated us to conduct a multi-year, high-cadence spectroscopic RM campaign of NGC 4151, an ideal target for investigating these issues. NGC 4151 is a well-studied Seyfert 1 galaxy that has exhibited CL behavior multiple times \citep{Li2022} throughout its observational history. Historical light curves spanning several decades reveal persistent large-amplitude variability, with the X-ray to optical flux varying by factors of several \citep{Couto2016}. These dramatic flux changes make NGC 4151 an exceptional laboratory for probing disk physics across varying accretion states.

\section{Data} \label{sec:2}
\subsection{Observations} \label{sec:2.1}
The data reported in this work are obtained in the 2024--2025 observing season, during which NGC 4151 entered its brightest state in nearly two decades. The optical RM campaign of NGC 4151 is initiated with the 2.4 m telescope at Lijiang Observatory, operated by Yunnan Observatories, Chinese Academy of Sciences (PI: H.-C. Feng), and includes both spectroscopic and photometric monitoring. The overall observational strategy and program design are described in detail in \citet{Feng2024}. During the 2022--2023 observing season, we have successfully measured inter-band continuum time delays using spectroscopic data alone, employing Grism 14 with a UV-blocking filter and a 5$\farcs$05 slit \citep{Zhou2025}.

To better constrain the shape of the DC component, we adopt a revised observing strategy in the current season. Each night, we obtain two spectra using Grism 14 (without the UV-blocking filter) and Grism 8, yielding complete wavelength coverage from 3600 to 9500 \AA. Simultaneous photometric observations are conducted in the $B$, $V$, $R$, and $I$ bands. The median cadences of the spectroscopic and photometric observations are $\sim$1.2 and $\sim$2.0 days, respectively. 

Motivated by the unexpectedly long lag from optical-only measurement reported in our previous work \citep{Zhou2025}, which significantly exceeds theoretical expectations, we also propose a new \textit{Swift} monitoring program to extend our wavelength coverage into the ultraviolet (UV) and X-ray regimes. When processing the data, we identify an extensive high-cadence \textit{Swift} dataset obtained by the ``Intensive Broadband Reverberation Mapping'' (IBRM) team, with daily sampling from MJD 60611--60871 \citep[e.g.,][]{Edelson2015,Edelson2017,Edelson2019,Cackett2020,Hernandez2020,Edelson2024}. This program (PI: R. Edelson\footnote{Swift Guest Investigator proposal number 2124077, \url{https://swift.gsfc.nasa.gov/proposals/c21_acceptarg.html}}
\footnote{\url{https://www.swift.psu.edu/toop/too_detail.php?id=21028}}) is part of a large multiwavelength campaign that also includes X-ray monitoring with \textit{NICER}, as well as ground-based optical and infrared photometry and spectroscopy (Partington et al., in preparation). Our own \textit{Swift} program, which spans MJD 60766--60856 (PI: C. Pinto\footnote{\url{https://www.swift.psu.edu/toop/too_detail.php?id=21851}}), overlaps with the IBRM dataset and, together with additional observations obtained during MJD 60645--60689 (PI: J. Miller\footnote{\url{https://www.swift.psu.edu/toop/too_detail.php?id=21573}}), provides sub-daily cadence during this interval.

Given the high brightness of the target, UVOT observing mode 0x224d, featuring a short readout time of 3.6 ms, is used to minimize saturation of the nuclear emission. Although our original request specifies mode 0x30d5, which has a longer readout time (11 ms), the configuration is updated at the start of the campaign to mode 0x224d, ensuring the acquisition of usable, unsaturated data.

\subsection{Data reduction} \label{sec:2.2}
The optical spectroscopic and photometric data from the Lijiang 2.4 m telescope were reduced following the procedures described in \citet{Feng2025}. In short, the reduction includes bias subtraction, flat correction, and wavelength calibration. Both photometric and spectroscopic fluxes were calibrated using comparison field stars observed simultaneously in the image or slit. To recover the absolute spectral shape and AGN fluxes, we applied an additional flux correction based on standard spectrophotometric stars observed under photometric conditions. Telluric absorption features in the spectra were removed using the comparison stars as reference.

The \textit{Swift} UVOT \citep{Roming2005} and XRT \citep{Burrows2005} data were processed following an approach similar to \citet{Edelson2017}, with minor modifications to account for the current properties of NGC 4151. Due to contamination from the host galaxy, we selected a background region free of host galaxy and field star emission for background subtraction in UVOT images. To minimize the impact of coincidence loss, we exclude all UVOT measurements with \texttt{FLUX\_AA} / \texttt{FLUX\_AA\_COI\_LIM} $\geqslant$ 1.0\footnote{\url{https://www.swift.ac.uk/analysis/uvot/}}, which indicate detector saturation. For the $U$ band, we adopt a more lenient limit of \texttt{FLUX\_AA} / \texttt{FLUX\_AA\_COI\_LIM} $\geqslant$ 1.4. We also removed data points flagged with \texttt{MAG} = 99, short exposures, or distorted PSFs.

XRT data were processed using the online \textit{Swift} XRT pipeline\footnote{\url{https://www.swift.ac.uk/user_objects/}}, using the ``snapshot" binning mode. Light curves were extracted in the 0.3-10 keV band, as well as in soft (0.3-1.5 keV) and hard (1.5-10 keV) bands to obtain the (X-ray) hardness ratio. To characterize the long-term variability of NGC 4151, we additionally analyzed all available \textit{Swift} archival data from 2016 to 2025 (Target IDs: 37258, 10655, 95669, 96883, 34455, and 15871). Nonetheless, the present study focuses exclusively on data obtained during the 2024--2025 observing season.

\begin{figure*}[ht!]
\includegraphics[scale=0.52]{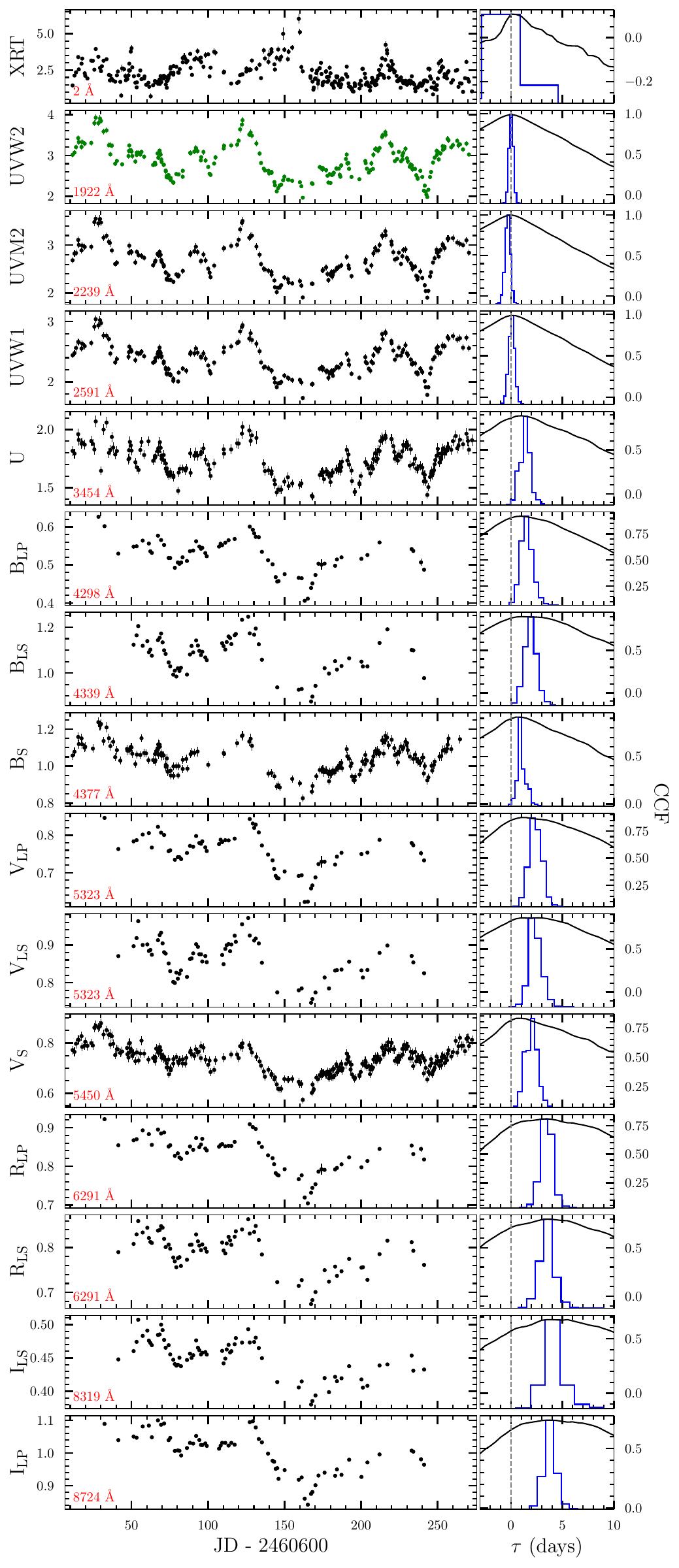}
\includegraphics[scale=0.52]{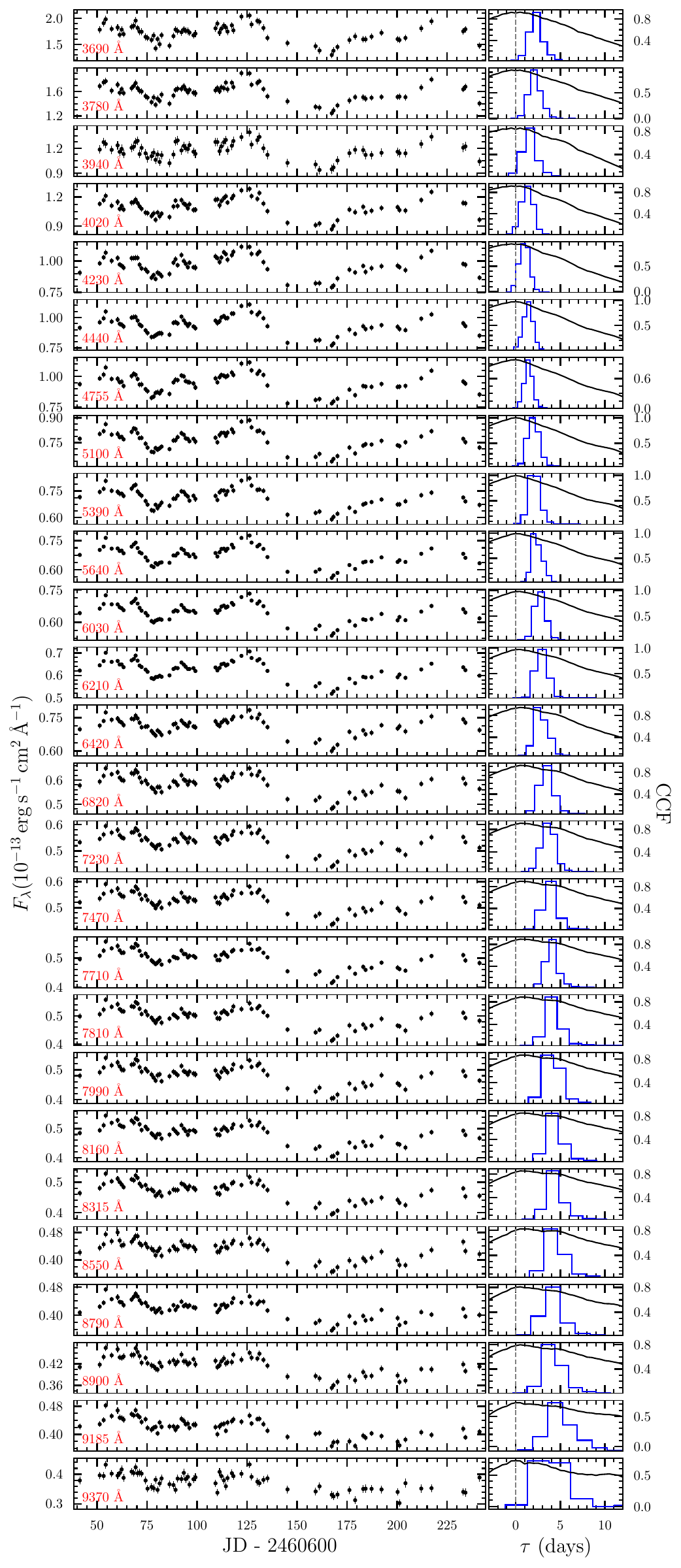}
\caption{Light curves and CCF analyses of NGC 4151. The left panel shows the photometric light curves from \textit{Swift} and Lijiang observations, where the subscript S represents \textit{Swift} data, LP represents Lijiang photometric data, and LS represents synthetic photometric data derived from Lijiang spectra. The right panel displays the light curves at different wavelengths measured from the line-free regions of the spectra. On the right side of each plot are the corresponding CCF and CCCD results.
\label{fig1}}
\end{figure*}

\section{Analysis} \label{sec:3}
\subsection{Multi-wavelength light curves}
Figure \ref{fig1} presents our intensive monitoring campaign of NGC 4151 during the 2024--2025 observing season, combining \textit{Swift} UV/X-ray observations with ground-based optical photometry and spectroscopy from the Lijiang 2.4 m telescope. The photometric light curves are obtained following the procedures described in Section \ref{sec:2.2}, while the spectroscopic light curves are extracted from carefully selected line-free regions. To verify the consistency between spectroscopic and photometric measurements, we convolve the Lijiang spectra with the transmission curves of the $B$, $V$, $R$, and $I$ filters. For each band, we compared the synthetic light curve with the corresponding direct photometric light curve. The resulting Pearson correlation coefficients exceed 0.93 in all bands, demonstrating excellent agreement and confirming the reliability of our spectroscopic flux calibration. The wavelength information for both filters and spectroscopic line-free regions is detailed in Table \ref{tab:1}, while the light curve data (including the long-term \textit{Swift} observations) are tabulated in Table \ref{tab:2}.

\begin{deluxetable}{lcccccccccc}[!ht]
\tablecolumns{10}
\tablewidth{\textwidth}
\tabletypesize{\scriptsize}
	\tablecaption{Summary of wavelengths and measured lags. \label{tab:1}}	
	\tablewidth{\textwidth}
		\tablehead{\multicolumn{1}{l}{Filter} &
			\colhead{$\lambda_{\rm phot}$} &	
            \colhead{Lag} &
			\colhead{Range} &	
            \colhead{$\lambda_{\rm spec}$} &	
			\colhead{Lag}       \\
            \colhead{} &
			\colhead{(\AA)} &	
            \colhead{(days)} &
			\colhead{(\AA)} &	
            \colhead{(\AA)} &	
			\colhead{(days)}       
		}
\startdata
UVW2 & 1922 & $0.00_{-0.24}^{+0.24}$ & 3680-3700 & 3690 & $2.21_{-0.71}^{+0.77}$\\
UVM2 & 2239 & $-0.25_{-0.25}^{+0.25}$ & 3770-3790 & 3780 & $2.02_{-0.56}^{+0.77}$\\
UVW1 & 2591 & $0.01_{-0.25}^{+0.24}$ & 3930-3950 & 3940 & $1.48_{-0.75}^{+0.77}$\\
$\rm U$ & 3454 & $1.25_{-0.48}^{+0.50}$ & 4010-4030 & 4020 & $1.24_{-0.52}^{+0.69}$\\
$\rm B_{LP}$ & 4298 & $1.49_{-0.49}^{+0.68}$ & 4220-4240 & 4230 & $0.98_{-0.51}^{+0.50}$\\
$\rm B_{LS}$ & 4339 & $1.77_{-0.53}^{+0.69}$ & 4430-4450 & 4440 & $1.43_{-0.65}^{+0.50}$\\
$\rm B_{S}$ & 4377 & $0.99_{-0.45}^{+0.47}$ & 4745-4765 & 4755 & $1.45_{-0.50}^{+0.49}$\\
$\rm V_{LP}$ & 5323 & $2.24_{-0.51}^{+0.69}$ & 5075-5125 & 5100 & $1.96_{-0.51}^{+0.54}$\\
$\rm V_{LS}$ & 5323 & $2.26_{-0.53}^{+0.70}$ & 5380-5400 & 5390 & $2.18_{-0.67}^{+0.54}$\\
$\rm V_{S}$ & 5450 & $1.94_{-0.49}^{+0.70}$ & 5630-5650 & 5640 & $2.43_{-0.64}^{+0.53}$\\
$\rm R_{LP}$ & 6291 & $3.44_{-0.53}^{+0.53}$ & 6020-6040 & 6030 & $2.71_{-0.51}^{+0.70}$\\
$\rm R_{LS}$ & 6291 & $3.49_{-0.73}^{+0.72}$ & 6200-6220 & 6210 & $2.92_{-0.67}^{+0.55}$\\
$\rm I_{LS}$ & 8319 & $4.02_{-0.74}^{+0.96}$ & 6410-6430 & 6420 & $2.75_{-0.70}^{+0.90}$\\
$\rm I_{LP}$ & 8724 & $3.71_{-0.52}^{+0.52}$ & 6810-6830 & 6820 & $3.46_{-0.74}^{+0.75}$\\
&  &  & 7220-7240 & 7230 & $3.68_{-0.74}^{+0.66}$\\
&  &  & 7460-7480 & 7470 & $3.89_{-0.76}^{+0.74}$\\
&  &  & 7700-7720 & 7710 & $3.98_{-0.60}^{+0.74}$\\
&  &  & 7800-7820 & 7810 & $4.13_{-0.69}^{+0.79}$\\
&  &  & 7980-8000 & 7990 & $4.01_{-0.78}^{+0.88}$\\
&  &  & 8150-8170 & 8160 & $4.26_{-0.72}^{+0.96}$\\
&  &  & 8300-8330 & 8315 & $4.25_{-0.76}^{+0.99}$\\
&  &  & 8540-8560 & 8550 & $4.25_{-0.80}^{+1.00}$\\
&  &  & 8780-8800 & 8790 & $3.98_{-0.94}^{+1.06}$\\
&  &  & 8885-8915 & 8900 & $4.05_{-0.84}^{+1.19}$\\
&  &  & 9170-9200 & 9185 & $4.74_{-1.11}^{+2.00}$\\
&  &  & 9340-9400 & 9370 & $3.91_{-1.35}^{+2.30}$\\
\enddata
\tablecomments{
$\lambda_{\rm phot}$ and $\lambda_{\rm spec}$ denote the central wavelengths of the filters and spectroscopic bins, respectively, both corrected to the rest frame. All lag values listed in the table have also been corrected to the rest frame. Subscripts indicate the data source: S = \textit{Swift}/UVOT, LP = Lijiang photometry, LS = Lijiang spectroscopy-derived synthetic photometry. Due to the limited spectral coverage, the $B_{\rm LS}$ and $I_{\rm LS}$ bands are only partially covered by the spectra; for these bands, $\lambda_{\rm phot}$ corresponds to the mean wavelength of the covered portion of the bandpass.
}
\end{deluxetable}

\begin{deluxetable*}{lcccccccccc}[!ht]
\tablecolumns{10}
\tablewidth{\textwidth}
\tabletypesize{\scriptsize}
	\tablecaption{Light curves from \textit{Swift} and Lijiang observations \label{tab:2}}	
	\tablewidth{\textwidth}
        \tablehead{\multicolumn{4}{c}{\textit{Swift}} &	&
            \multicolumn{3}{c}{Lijiang}\\
            \cline{1-4} \cline{6-8}
            \multicolumn{1}{l}{JD} &
            \colhead{Flux} &            
			\colhead{\texttt{frametime}} &	
            \colhead{Filter} &&            
            \colhead{JD} &	
            \colhead{Flux} &
            \colhead{Range/Filter} 
		}
\startdata
2453730.9825 & $0.532\pm 0.012$  &  11.0 &UVW2 && 2460651.4062 & $1.784\pm 0.066$  & 3680-3700 \\
2453731.0506 & $0.529\pm 0.011$  &  11.0 &UVW2 && 2460653.3617 & $1.908\pm 0.065$  & 3680-3700 \\
2453731.1182 & $0.530\pm 0.011$  &  11.0 &UVW2 && 2460654.3975 & $1.983\pm 0.066$  & 3680-3700 \\
2453731.1860 & $0.530\pm 0.011$  &  11.0 &UVW2 && 2460657.3648 & $1.798\pm 0.065$  & 3680-3700 \\
2453731.2528 & $0.536\pm 0.011$  &  11.0 &UVW2 && 2460660.3520 & $1.826\pm 0.065$  & 3680-3700 \\
\enddata
\tablecomments{Both photometric and spectroscopic data have been converted to flux. All fluxes are in units of $10^{-13}~\ergscma$, except for the Lijiang photometric data, which are in arbitrary flux units. The frametime is in ms.
\\
(This table is available in a machine-readable form in the online journal.)}
\end{deluxetable*}

\begin{figure*}[ht!]
\includegraphics[scale=0.45]{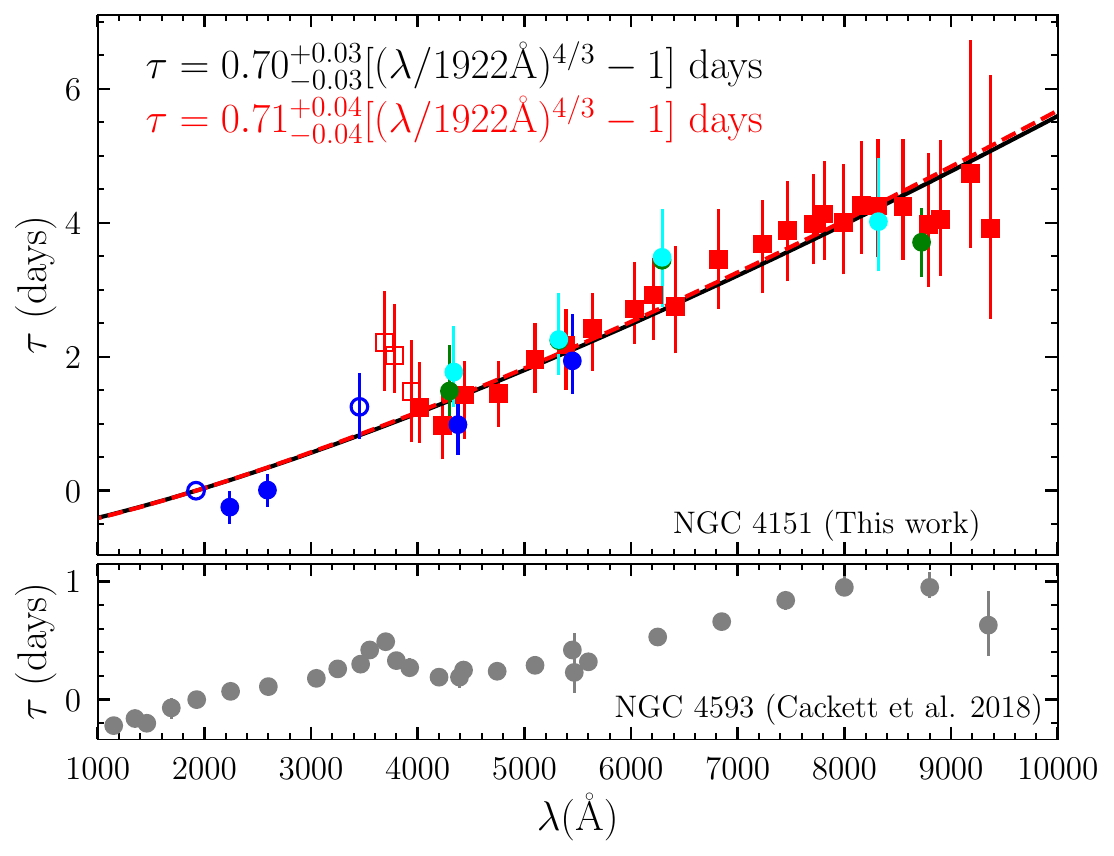}
\includegraphics[scale=0.45]{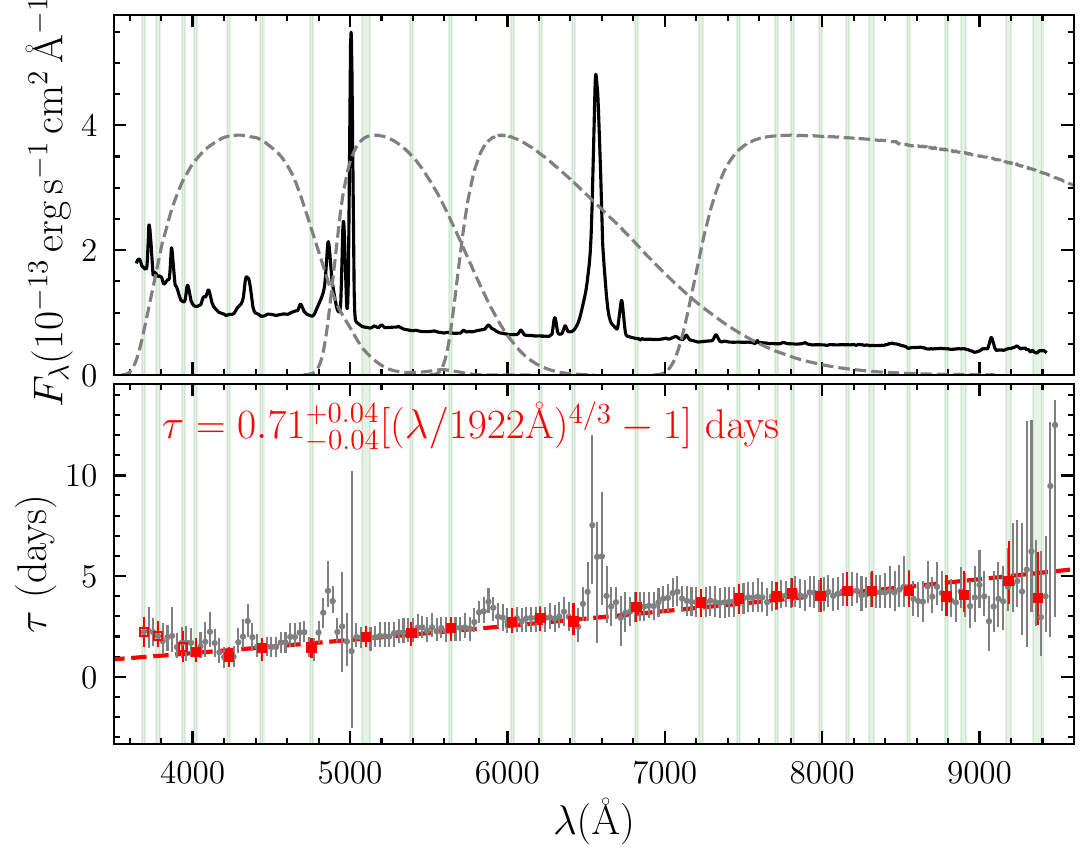}
\caption{Wavelength-dependent lags and test of the line-free region. Left: The top panel shows lags for NGC 4151 derived from \textit{Swift} photometry (blue dots), Lijiang photometry (green dots), synthetic photometry (cyan dots), and spectroscopy (red squares). The black solid line and red dashed line represent fits to $\tau = \tau_{0}[(\lambda / 1922 {\rm \AA})^{4/3} - 1]$, using all data points and spectroscopy-only data, respectively. Hollow points were excluded from the fits. The bottom panel shows lags for NGC 4593 from \citet{Cackett2018}. Right: The mean spectrum with Lijiang filter transmission curves (top) and the lags measured at different wavelengths (bottom). The spectrum and transmission have been corrected to the rest frame. Gray dots indicate lags measured from the spectrum at every 30 \AA, while the red squares and red dashed line are the same as those in the left panel. The green shaded regions mark the line-free wavelength ranges chosen for the analysis.
\label{fig2}}
\end{figure*}

\subsection{Wavelength-dependent lags}
We measure the inter-band time lags using the interpolated cross-correlation function \citep[CCF;][]{Gaskell1987}, adopting the UVW2 band as reference. The associated uncertainties were estimated via flux randomization and random subset sampling (RF/RSS) to generate cross-correlation centroid distributions \citep[CCCD;][]{Peterson1998, Peterson2004}. Because the light curves from different bands have unequal cadences, a simple linear interpolation on the more sparsely sampled curve to match the denser one may introduce bias and increased uncertainties. We implemented an adaptive interpolation scheme: for each light curve pair, we interpolate the lower-cadence curve onto the time grid of the higher-cadence one, thereby minimizing interpolation effects.

The CCFs and CCCD histograms are shown in Figure~\ref{fig1}, and the measured lags and their $1\sigma$ uncertainties are listed in Table \ref{tab:1} (positive lags indicate a delayed response with respect to the UVW2 band). We notice that the CCF of spectroscopic data tend to peak at nearly-zero lag, which may be caused by flux calibration issues. For the spectra, systematic flux offsets arising from telescope tracking errors or slit miscentering will simultaneously affect flux measurements across all wavelength bins. These fluctuations can artificially enhance the CCF at zero lag. To test this hypothesis, we also apply the FR/RSS method to generate cross-correlation peak distributions (CCPDs). The RSS can randomly exclude ``outlier" data points, while the FR will randomly redistribute their flux values. Therefore, CCPDs are expected to reveal a shift toward longer lags that better represent the true physical delays. Indeed, our resulting CCPDs show this systematic shift toward longer lags. In this work, we adopt centroid lag measurements to further mitigate the influence of these outliers.

Although UV and optical variability is often interpreted as being driven by reprocessing of X-ray emission \citep[see][for a review]{Cackett2021}, our \textit{Swift}/XRT observations do not show any significant correlation with the UV or optical light curves. This lack of correlation, which is common in AGNs \citep{Kammoun2024}, is reflected in a low maximum CCF coefficient of $r_{\rm max} \sim 0.1$ (Figure \ref{fig1}). We therefore exclude the X-ray data from our lag analysis and restrict it to the UVOT and Lijiang bands.

Figure \ref{fig2} (left panel) displays the measured time lags as a function of wavelength. The lags increase systematically from the UV to the optical, consistent with expectations for an inside out propagation process. We fit the lag-wavelength relation using $\tau = \tau_{0}[(\lambda / \lambda_{0})^{4/3} - 1]$, where $\lambda_{0} = 1922$ \AA\ corresponds to the effective rest-frame wavelength of the UVW2 band. The $\lambda^{4/3}$ dependence arises from the temperature profile of a standard thin disk ($T \propto r^{-3/4}$) combined with Wien's law, which implies that emission at wavelength $\lambda$ originates predominantly from radius $r \propto \lambda^{4/3}$.

Several data points are excluded from the fit. The \textit{Swift}/$U$ band is omitted because the source brightness exceeded the \texttt{FLUX\_AA\_COI\_LIM} threshold for most observations, resulting in unreliable photometry. This region may also be contaminated by strong Balmer continuum emission \citep{Korista2019}. Accordingly, spectroscopic points in the same wavelength range are excluded as well. A fit to the spectroscopic data alone yields $\tau_0 = 0.71 \pm 0.04$ days. Including both spectroscopic and photometric data results in a consistent value of $\tau_0 = 0.70 \pm 0.03$ days.

Given the cadence mismatch between Lijiang and \textit{Swift} data, and motivated by studies suggesting that lag measurements can be cadence-dependent \citep[e.g.,][]{Vincentelli2021, Cackett2022, Chen2024}, we perform an additional consistency check by interpolating the 5100 \AA\ light curve onto the time grids of other bands. Fitting the relation $\tau = \tau_{5100}[(\lambda / 5100 {\rm \AA})^{4/3} - 1]$ yielded $\tau_{5100} = 2.59 \pm 0.30$ days for spectroscopic data only and $\tau_{5100} = 2.68 \pm 0.24$ days for the combined dataset (Appendix \ref{sec:AppxA}). These correspond to $\tau_{0} = 0.70 \pm 0.08$ and $0.73 \pm 0.07$ days, respectively, when scaled to 1922 \AA, in excellent agreement with our main results. 

We also analyze the structure function of each light curve (Fu et al. in preparation), and find that variability of \textit{Swift} on timescales shorter than 1 day may not originate from intrinsic AGN variability. Although space-based instruments typically maintain stable PSFs, our tests show that changes in aperture size introduce significant noise in flux measurements, indicating that the host galaxy can indeed affect variability measurements (Appendix \ref{sec:AppxB}). Therefore, our aperture photometry measurements of \textit{Swift} data can only reliably trace AGN variations on timescales longer than 1 day, which substantially mitigates the cadence difference between ground-based and space-based observations. Nevertheless, examination of the light curve profiles confirms that these effects do not substantially bias the lag measurement, which remains dominated by variability on multi-day timescales.

To validate our choice of line-free regions, we compute CCFs for the entire spectrum in 30 \AA\ intervals. The resulting lag spectrum confirms that our selected regions are minimally affected by emission lines and follow the continuum lag relation well (the right panel of Figure \ref{fig2}).

\begin{figure*}[ht!]
\centering 
\includegraphics[scale=0.65]{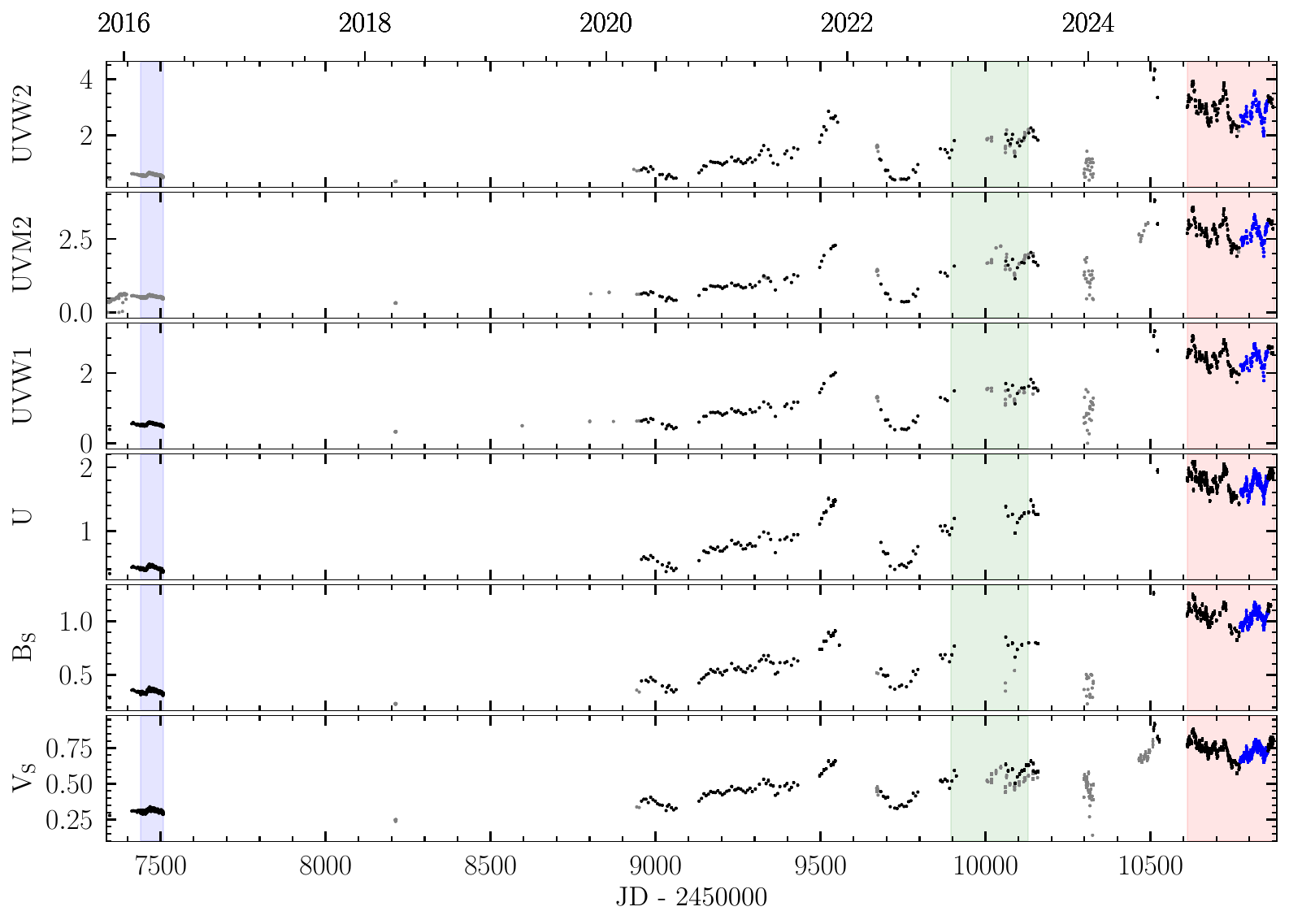}
\caption{Long-term \textit{Swift}/UVOT light curve of NGC 4151. The black points represent observations with a frametime of 3.6 ms, while gray points correspond to observations with an 11 ms frametime. The blue points highlight the observation period of our monitoring program. The purple, green, and red shaded regions indicate the durations of three RM campaigns conducted by \citet{Edelson2017}, \citet{Zhou2025}, and this work, respectively.
\label{fig3}
}
\end{figure*}

\subsection{Evolution of lags with luminosity} \label{sec:3.3}
Figure \ref{fig3} presents the decade-long \textit{Swift}/UVOT light curves of NGC 4151, revealing dramatic variability over this period. Including our current campaign, three high-cadence continuum RM studies have now been conducted, allowing us to investigate how the continuum lags evolve with luminosity. We adopt the UVW2 band flux as a tracer of the intrinsic AGN variability, as this band is minimally affected by host-galaxy starlight contamination in UV/optical bands.

\citet{Edelson2017} reported results from an intensive \textit{Swift} monitoring campaign in 2016, when NGC 4151 was in a historically low luminosity state. They measured a lag of $\tau_{5100} = 1.25 \pm 0.40$ days.\footnote{ For consistent comparison, all $\tau_0$ values are scaled to represent the rest-frame lag of 5100 \AA.} During 2022--2023, the source became brighter by a factor of $\sim$3 compared to 2016. Our Lijiang 2.4 m spectroscopic monitoring revealed that the lag increased by a factor of 3.8 (4.7$\sigma$), reaching $\tau_{5100} = 4.78 \pm 0.63$ days \citep{Zhou2025}. In our current campaign, NGC 4151 further brightened by a factor of 1.6, but the measured lag decreased to $\tau_{5100} = 2.59 \pm 0.30$ days.

To compare our measured lags with theoretical expectations, we adopt the analytical prescription of the X-ray-reprocessing standard thin disk model \citep[e.g.,][]{Krolik1991, Cackett2007}. In this model, the lag $\tau$ at wavelength $\lambda$ scales as $\tau(\lambda) \propto \lambda^{4/3}(M_{\rm BH} \dot{M})^{1/3}$, where $M_{\rm BH}$ is the black hole mass and $\dot{M}$ is the mass accretion rate. Following \citet{Fausnaugh2016} and \citet{Zhou2025}, the wavelength-dependent lag can be expressed as:
\begin{equation} \label{eq1}
\tau(\lambda) = \frac{2.49^{4/3}}{c} \left( \frac{k \lambda}{hc} \right)^{4/3}
\left( \frac{3(1 + k_X) G M_{\rm BH} \dot{M}}{8\pi \sigma_{\text{SB}}} \right)^{1/3},
\end{equation}
where $k_X = 1/3$, adopted following \citet{Zhou2025}, is the ratio of external X-ray heating to internal viscous heating, $G$ is the gravitational constant, $c$ is the speed of light, $h$ is the Planck constant, $k$ is the Boltzmann constant, and $\sigma_{\text{SB}}$ is the Stefan-Boltzmann constant.

\begin{figure}[ht!]
\centering 
\includegraphics[scale=0.55]{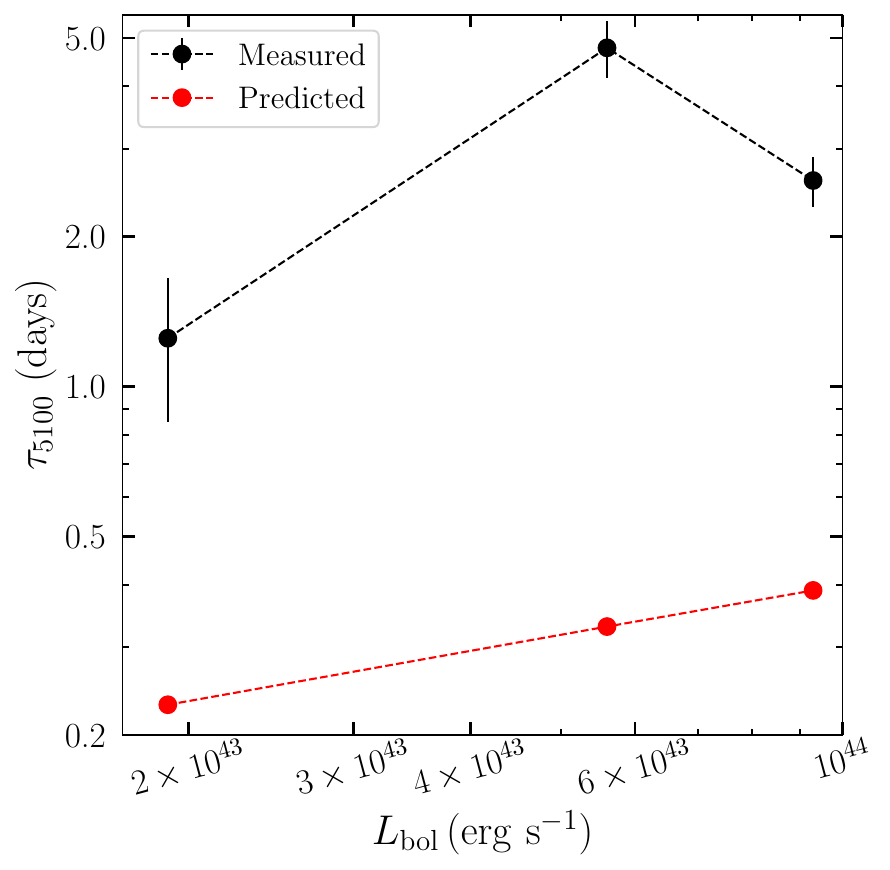}
\caption{Lag vs. luminosity for the three campaigns. Black and red represent the measured time delays and the predictions from X-ray reprocessing in a standard thin accretion disk model, respectively.
\label{fig4}}
\end{figure}

We adopt a black hole mass of $M_{\rm BH} = 4.27 \times 10^7 M_{\odot}$ from stellar dynamical modeling \citep{Onken2014, Roberts2021}. The accretion rate is inferred from the bolometric luminosity, which is typically estimated from monochromatic measurements. However, differences in methodology and data quality across studies can introduce uncertainties. To minimize such inconsistencies, we adopt the bolometric luminosity $L_{\rm bol} = 5.6 \times 10^{43}$ erg s$^{-1}$ from \citet{Zhou2025} as our reference. This value was derived by subtracting the host-galaxy contribution using HST image decomposition. Based on the relative UVW2 fluxes, we infer bolometric luminosities of $1.9 \times 10^{43}$ and $9.2 \times 10^{43}$ erg s$^{-1}$ for the other two RM studies.\footnote{The dominant uncertainty arises from the adopted $ L_{\rm bol}$ measurement, but this introduces only a systematic offset in the comparison between observed and predicted lags, and does not affect our analysis of lag-luminosity trends.} The resulting theoretical lag predictions at 5100 \AA\ are 0.23, 0.33, and 0.39 days for the three RM campaigns, while the observed values are larger by factors of 5.4, 14.5, and 6.6, respectively. The measured and predicted variations of $\tau_{5100}$ with luminosity over the three campaigns are shown in Figure \ref{fig4}.

\section{Results and Discussion} \label{sec:4}
Our intensive spectroscopic and photometric RM campaign enables us to measure inter-band continuum lags in NGC 4151 from the UV to optical wavelengths. The use of spectroscopy allows us to remove contamination from broad emission lines and to construct a more detailed profile of the DC component. Time delays measured from line-free spectral regions follow a tight $\tau \propto \lambda^{4/3}$ relation, consistent with reprocessing in a stratified disk, and align well with the UV lags derived from \textit{Swift}/UVOT photometry. However, the measured lags are significantly larger than predicted by the standard X-ray-reprocessing thin-disk model by a factor of 6.6. Moreover, we find negligible correlation between X-ray and UV/optical variations for most of the monitoring period, challenging the framework in which X-ray irradiation is the dominant driver of disk variability.

The lack of X-ray/optical correlation is not unique to NGC 4151. A similar behavior has been reported in other AGNs \citep[e.g.,][]{McHardy2018, Edelson2019}, suggesting that the simple picture of X-ray reprocessing is incomplete. Several mechanisms could explain this behavior: (1) The X-ray corona may be geometrically decoupled from the disk, either through changes in the height or radial extent of the X-ray corona, which modify the covering fraction and hence the solid angle subtended by the disk \citep{Wilkins2017, Alston2020, XrismCollaboration2024}; (2) The observed X-rays may not represent the illuminating flux due to anisotropic emission or variable obscuration along our line of sight \citep{Gaskell2023, Panagiotou2022, Xiang2025, Lewin2025}; (3) Additional heating mechanisms such as magnetic dissipation or local accretion energy release may dominate the UV/optical emission \citep{Ren2024b, Secunda2024}, reducing the fractional contribution from X-ray reprocessing.

Our spectroscopic lag spectrum reveals clear excess delays near the Balmer jump (3647 \AA) and possibly near the Paschen jump (8206 \AA). The shape of the lag spectrum closely resembles that observed in NGC 4593 \citep{Cackett2018}, where DC contamination has been well-established. Given that DC originates from a more extended region than the disk, its contribution introduces a time-delayed signal that lies on top of the intrinsic disk response. Therefore, (at least part of) the enhanced lags we observe can be attributed to the DC component.

If the DC contribution dominates the lag signal, one would expect that the observed lag increases with luminosity, as the DC-emitting region expands with increased photoionizing flux \citep[just like broad lines,][]{Korista2001}. This expectation is broadly satisfied when comparing our current dataset to the 2016 low-luminosity campaign \citep{Edelson2017}: the UVW2 flux increased by a factor of 4.8, while the measured lag increased by a factor of 2.1, consistent with the scaling $\tau \propto L_{\rm bol}^{1/2}$. However, this trend breaks down when comparing our campaign with the 2022--2023 season. Despite a 1.6-fold increase in UVW2 flux, the continuum lag decreased by a factor of 1.8 (3.1$\sigma$). We stress that the H$\beta$ lag of the current campaign is also larger than that of the 2022--2023 season by a factor of two (Feng et al. in preparation), indicating that the emission regions of BLR clouds indeed expanded from the 2022--2023 season to the current season. This non-monotonic behavior (as shown in Figure \ref{fig4}) challenges the conventional DC model, which predicts a monotonic increase in lag with luminosity \citep{Netzer2022} due to the expansion of the DC-emitting region under stronger photoionizing flux.

This puzzling behavior may be understood if the measured lag represents a luminosity- or responsivity-weighted average radius from multiple regions. As the AGN brightens, the disk emission may increase more significantly than the DC component, thereby shifting the weighted mean lag toward smaller values. This suggests that the DC component itself also follows an intrinsic Baldwin effect \citep{Pogge1992}, with its relative contribution diminishing as the continuum luminosity increases. Such behavior is further supported by statistical trends observed across AGN populations, where more luminous sources tend to exhibit shorter continuum lags \citep{Li2021}. The particularly short lag observed during the low-luminosity state in 2016 may reflect compact DC and disk emission regions at such faint levels, resulting in short light-travel times across all reprocessing zones.

Alternatively, the anomalous lag behavior may reflect intrinsic properties of the accretion system itself. One possibility is structural evolution of the accretion disk, as expected in CL AGNs, where changes in disk geometry or radiative efficiency may lead to lag variations that deviate from the predictions of standard disk models. Indeed, the dramatic variability of NGC 4151 is accompanied by recurrent CL behavior, with historical transitions between Type 1.5 and 1.8 \citep{shapovalova2008}, and even a near Type 2 state in 1984 \citep{Penston1984}. Notably, during the period from 2016 to 2023, it underwent another spectral change from Type 1.8 to Type 1.5 \citep{Zhou2025}, potentially lending support to this interpretation. This scenario may be further examined using the Eddington ratios estimated from the SMBH mass and bolometric luminosities in Section \ref{sec:3.3}, where the Eddington luminosity is given by $L_{\rm Edd} = 1.26 \times 10^{38} M_{\rm BH}$/\msun\ erg s$^{-1}$. The resulting accretion rates are 0.0035, 0.0104, and 0.0170 for the three RM campaigns. These values span the theoretical accretion rate threshold at an Eddington ratio of $\sim$0.01, where a transition from geometrically thick to thin-disk accretion is expected to occur \citep{Narayan1995}.

Another possibility involves magnetohydrodynamic processes: magnetic perturbations in the corona can drive X-ray variability, triggering temperature fluctuations in the disk. As these fluctuations propagate outward, the disk heating rate varies accordingly and drives UV/optical variations on thermal timescales \citep{Sun2020}. While longer lags are generally expected at higher luminosities due to increasing thermal timescales, inner-disk perturbations can induce faster responses, leading to the non-monotonic lag-luminosity behavior seen in NGC 4151 \citep{Zhou2025}.

\section{Summary}
Thanks to the high spectral resolution and the pronounced variability of NGC 4151, we are able to detect a robust ($>3\sigma$) non-monotonic relationship between the continuum lag and luminosity. To the best of our knowledge, this is the first detection of such phenomenon in CL AGNs. This complex lag behavior not only provides new insights into the structure and reprocessing efficiency of the accretion disk, but also has important implications for the size of the BLR. Previous studies have shown that time delays between the UV and optical continua can lead to a systematic underestimation of the BLR reverberation signal \citep{Li2022, Feng2024}, which in turn propagates into increased uncertainties in the measurement of SMBH masses.

These findings present fundamental challenges to the standard X-ray reprocessing paradigm for AGN accretion disks and offer new observational constraints on the structure and evolution of AGN:
\begin{enumerate}
\item Despite the excellent $\tau \propto \lambda^{4/3}$ scaling consistent with X-ray reprocessing from a standard thin disk, our spectroscopic lag spectrum reveals clear excess delays near the Balmer and possibly Paschen jumps. This demonstrates that the DC emission contributes a considerable fraction to the measured lags, implying that simple reprocessing models alone cannot fully account for the observed lag spectrum.

\item Combining three seasons of continuum RM spanning nearly a decade, we discover that continuum lags do not increase monotonically with luminosity. This non-monotonic behavior can be interpreted in the framework of an intrinsic Baldwin effect associated with the DC component, where its relative contribution decreases at higher luminosities, as well as potential structural evolution of the accretion disk. The latter possibility is particularly relevant in the context of CL AGNs, where the large variability amplitude is often attributed to changes in accretion structure.

\item Our observations provide crucial constraints on AGN structure across multiple scales. The complex correlation between disk and DC emission requires more sophisticated models for interpreting RM results and understanding their physical origins. Furthermore, the variable continuum lags can systematically bias BLR reverberation measurements, increasing uncertainties in SMBH mass determinations.
\end{enumerate}
Altogether, these results highlight the importance of long-term, high-cadence, broad-wavelength spectroscopic monitoring campaigns in disentangling the multi-component and evolving nature of AGN variability. While NGC 4151 offers a robust evidence for non-monotonic lag-luminosity behavior in AGNs, extending this analysis to a broader sample is essential to assess the generality of this phenomenon and to probe its underlying physical mechanisms. Given that the current findings are based on only three seasons of monitoring, future efforts will require extended, multi-year spectroscopic campaigns targeting a diverse population of AGNs.

\begin{acknowledgments}
We are grateful to the anonymous referee for helpful comments that improved the manuscript. We also thank Kim Page for her patient and detailed assistance with the \textit{Swift} data reduction and analysis. This work is supported by National Key R\&D Program of China (No. 2021YFA1600404 and 2023YFA1607903), the National Natural Science Foundation of China (Nos. 12203096, 12573021, 12303022, 12322303, and 12373018), Yunnan Fundamental Research Projects (grants NO. 202301AT070358 and 202301AT070339), the Fundamental Research Funds for the Central Universities (0620ZK1270), and the science research grants from the China Manned Space Project with (Nos. CMS-CSST-2025-A02, CMS-CSST-2025-A07). CP is supported by European Union - Next Generation EU, Mission 4 Component 1 CUP C53D23001330006. NK acknowledge  funds from  European  Union - Next Generation EU, Mission 4 Component 1 CUP C53D23001330006 and  INAF Large Grant 2023 BLOSSOM FO 1.05.23.01.13.

We acknowledge the support of the staff of the Lijiang 2.4 m telescope. Funding for the telescope has been provided by Chinese Academy of Sciences and the People’s Government of Yunnan Province. We acknowledge the use of public data from the \textit{Swift} data archive. We acknowledge the support of the staff from BOOTES telescopes.

\end{acknowledgments}





%
\facilities{YAO:2.4m, \textit{Swift}}

\software{Astropy \citep{Astropy2013}, HEASoft-6.31.1 \citep{HEASoft2014}, PYCCF \citep{Sun2018}, PyRAF \citep{Pyraf2012}}


\appendix
\renewcommand{\thefigure}{\Alph{section}\arabic{figure}}
\setcounter{figure}{0}
\section{Cadence-Matched lag measurements} \label{sec:AppxA}
\begin{figure*}[ht!]
\centering
\includegraphics[scale=0.45]{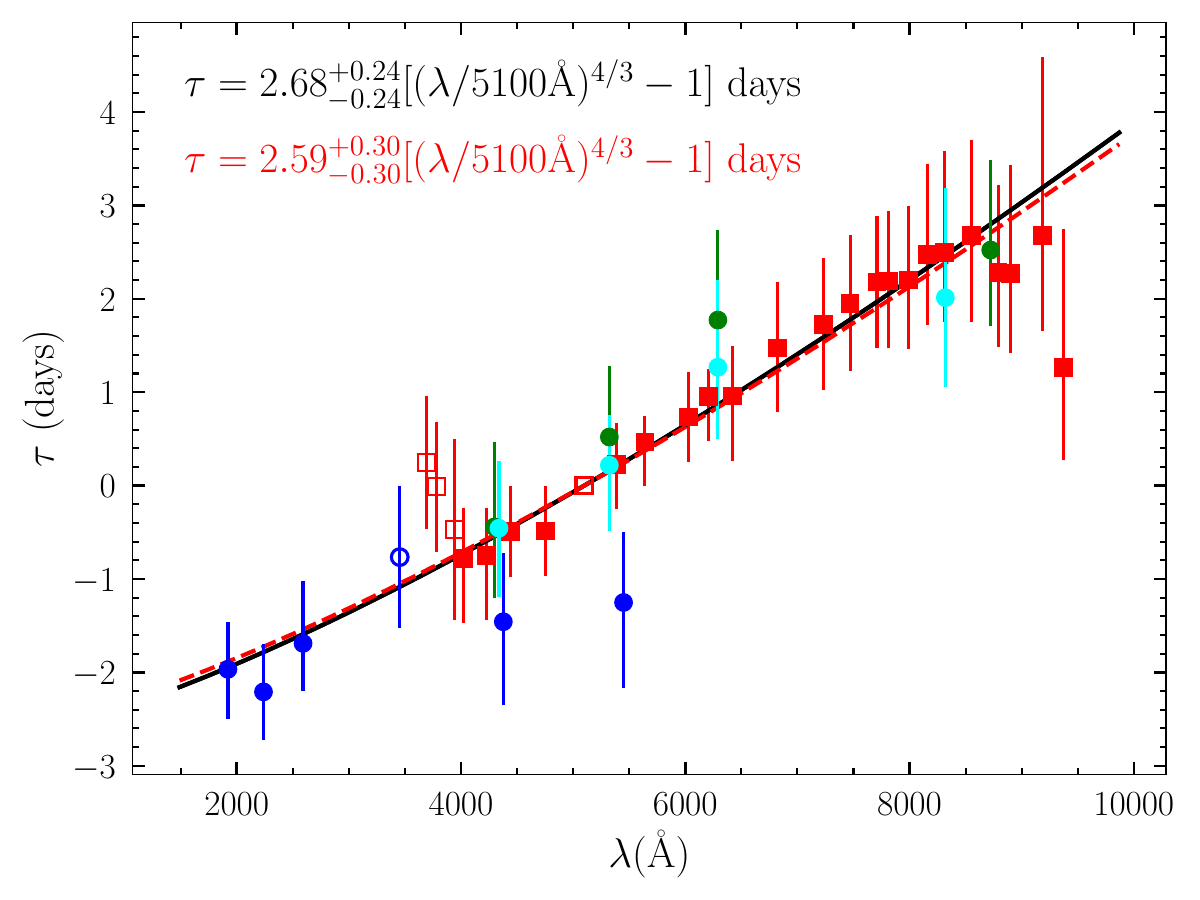}
\caption{Same as Figure \ref{fig2}, but with the CCFs calculated using the 5100 \AA\ light curve as the reference, and all other light curves interpolated to match the cadence of the 5100 \AA\ data.
\label{figa1}}
\end{figure*}

To assess the potential impact of cadence differences between telescopes on the lag measurements, we performed a consistency test using the Lijiang 5100 \AA\ light curve as the reference for CCF calculations. Since the Lijiang data have uniformly lower and comparable cadences across all bands, we interpolated all light curves to match the 5100 \AA\ time grid, ensuring that all lag measurements were derived from light curves with identical cadence. We then fitted the resulting lags using the relation $\tau = \tau_{5100}[(\lambda / 5100 {\rm \AA})^{4/3} - 1]$. The results are shown in Figure \ref{figa1}.  The overall wavelength-dependent trend remains consistent with the main analysis, suggesting that cadence mismatch does not significantly bias the derived lag-wavelength relation.

\setcounter{figure}{0}
\section{Testing the effect of aperture sizes on \textit{Swift}} \label{sec:AppxB}
\begin{figure*}[ht!]
\centering 
\includegraphics[scale=0.65]{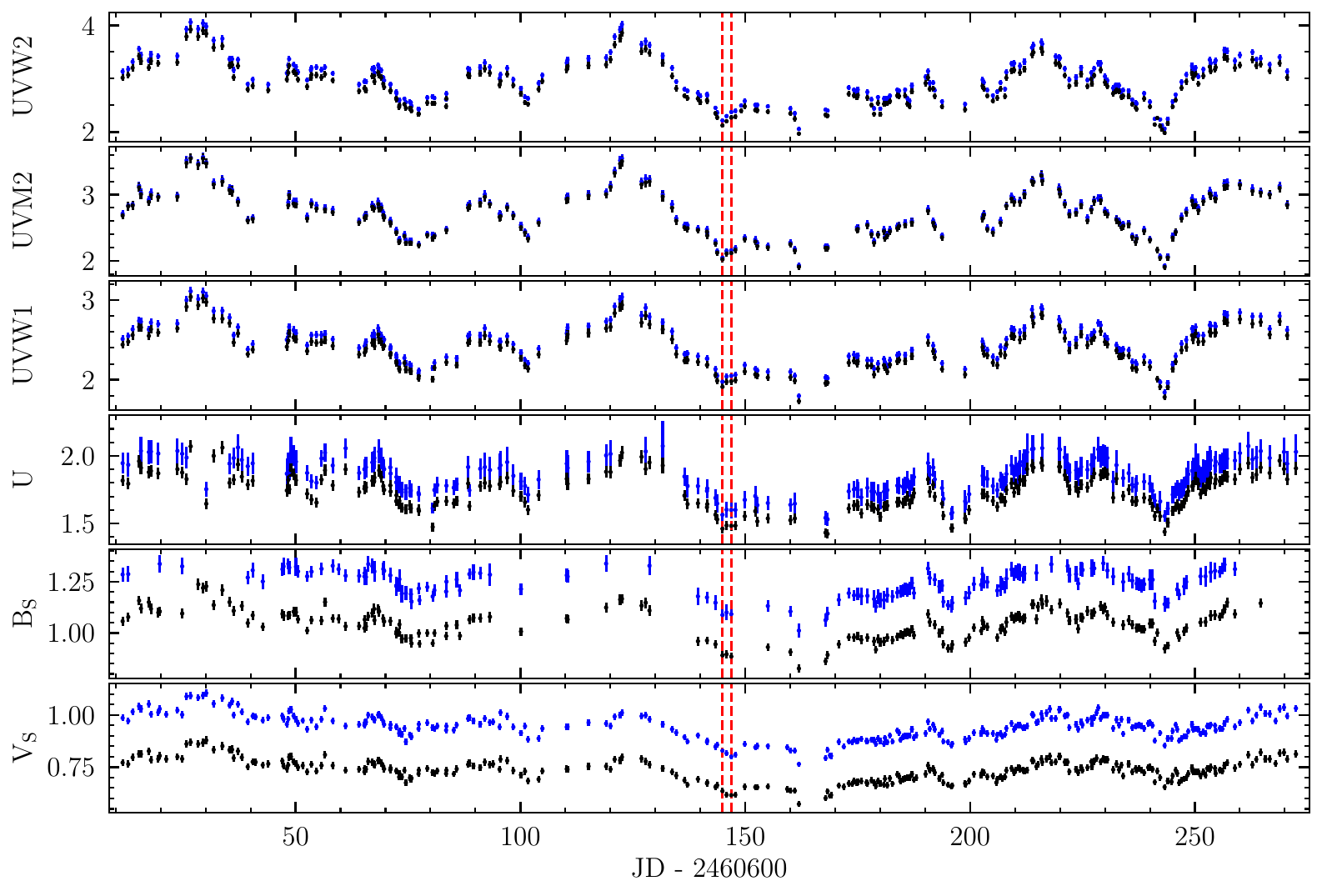}
\caption{Example of \textit{Swift}/UVOT light curves extracted with different apertures. In each panel, black and blue points represent aperture radii of 5\arcsec and 10\arcsec, respectively. The red dashed lines indicate a valley of UVW2 and the corresponding $V$-band response delayed by the measured (observed-frame) lag.
\label{figb1}}
\end{figure*}

We tested the potential impact of aperture size on our \textit{Swift} photometric measurements by extracting light curves using multiple aperture radii. In the UV bands, where host galaxy contribution is negligible, the light curves were nearly insensitive to aperture size. In contrast, for the optical bands---where the host contamination is stronger---larger apertures introduce noticeable uncertainties and increased scatter in the photometry. Similar differences between ground-based and \textit{Swift} observations, particularly in the optical-IR bands, have also been reported by \citet{Gonzalez-Buitrago2025}, and may be partially related to host galaxy contamination. Despite these differences, the overall variability trends remain consistent across apertures, and the inter-band lag measurements remain largely consistent. As an example, Figure \ref{figb1} shows the light curves extracted with 5\arcsec\ and 10\arcsec\ radius apertures. While the optical light curves exhibit increased noise with larger aperture, the delayed structure remains clearly visible.

{}

\end{document}